\documentstyle[11pt,newpasp,twoside]{article}
\def\edcomment#1{\iffalse\marginpar{\raggedright\sl#1\/}\else\relax\fi}
\reversemarginpar

\begin{document}
\title{Galactic Winds: Near and Far}
 \author{Sylvain Veilleux\altaffilmark{1,2}}
\affil{Department of Astronomy, University of Maryland, College Park, MD 20742}

\altaffiltext{1}{Current Address: 320-47 Downs Lab., Caltech, Pasadena,
CA 91125 and Observatories of the Carnegie Institution of Washington,
813 Santa Barbara Street, Pasadena, CA 91101}

\altaffiltext{2}{Cottrell Scholar of the Research Corporation}

\begin{abstract}
Multiwavelength data on star-forming galaxies provide strong evidence
for large-scale galactic winds in both nearby and distant objects. The
results from recent ground-based and space-borne programs are
reviewed. The impact of these winds on the host galaxies and the
surrounding environment is discussed in the context of galaxy
evolution.
\end{abstract}

\section{Introduction}

Galactic winds that extend on a scale comparable to the host galaxies
are now known to be a common feature both locally and at high
redshifts.  These winds are significant carriers of mass, momentum,
and energies that may impact the formation and evolution of the host
galaxies and the surrounding intergalactic medium. Given the scope of
this conference, the present review focusses exclusively on
starburst-driven winds. AGN-driven galactic winds, perhaps a very important
phenomenon in the lives of galaxies with spheroids (Kormendy \&
Gebhardt 2001), are not discussed here (see, e.g., Veilleux et
al. 2002a for a recent review of this topic). Due to space
limitations, the emphasis of this review is on the recent ($\ga$ 1998)
literature. Readers interested in results from earlier studies may
refer to the reviews by Strickland (2002) and Heckman (2002).

First, the basic physics of starburst-driven winds is described
briefly in \S 2.  An observational summary of the properties of local
winds is given in the preamble to \S 3. The remainder of \S 3
describes detailed data on three well-studied cases of local
starburst-driven winds, and summarizes the evidence for winds in
luminous and ultraluminous infrared galaxies and distant Lyman break
galaxies. This section often emphasizes the importance of using
multiwavelength data to draw a complete picture of this complex
multi-phase phenomenon.  The impact of starburst-driven winds on
the host galaxies and their environment is discussed briefly in \S
4. Here the focus of the discussion is on the existence and properties
of the wind fluid and on the size of the ``zone of influence'' of
these winds. A summary is given in \S 5.
%summarizes some of the evidence collected so far that suggests that
%galactic winds have had a strong impact on galaxy evolution and the
%environment.

\section{Basic Physics}

The driving force behind starburst-driven winds is the mechanical
energy from stellar winds and supernova events (e.g., Chevalier \&
Clegg 1985). This mechanical energy is quickly thermalized to produce
a hot cavity with a temperature $\sim 10^8\;\Lambda^{-1}$ K, where
$\Lambda = M_{\mathrm total}/M_{\mathrm ejecta} \ge 1$ is the
mass-loading term. This over-pressured cavity expands through the
ambient medium, sweeping this material up in the process to produce a
bubble-like structure.  The complex interaction between the wind and
the ISM of the host galaxy has been the subject of several numerical
simulations (e.g., MacLow \& McCray 1988; Suchkov et al. 1994, 1996;
MacLow \& Ferrara 1999; D'Ercole \& Brighenti 1999; Strickland \&
Stevens 2000; Silich \& Tenorio-Tagle 2001). If radiative energy
losses are negligible (probably a good assumption in some objects;
e.g., Heckman et al. 2001), the bubble expands adiabatically through
the galaxy ISM with a velocity $\sim
100~n_0^{-0.2}\;\dot{E}_{42}^{0.2}\;t_7^{-0.4}$ km s$^{-1}$, where
$n_0$ is the ambient nucleon density in cm$^{-3}$, $\dot{E}_{42}$ is
the rate of deposition of mechanical energy in 10$^{42}$ erg s$^{-1}$,
and $t_7$ is the age of the bubble in 10$^7$ years (e.g., Weaver et
al. 1977).

A powerful starburst may inject enough energy to produce a cavity of
hot gas that can burst out of the disk ISM,
%with dimensions comparable to the scale height of the disk ISM,
at which point the dense walls of the bubble start accelerating
outward, become Rayleigh-Taylor unstable, and break up into cloudlets
and filaments. If halo drag is negligible (probably {\em not} a good
assumption in general), the wind fluid may reach terminal velocities
as high as $\sim$ 3000 $\Lambda^{-1}$ km s$^{-1}$, well in excess of
the escape velocity of the host galaxy. In contrast, the terminal
velocities of clouds accelerated by the wind are more modest, of order
$\sim 600~\dot{p}_{34}^{0.5}\;\Omega_w^{-0.5}\;r_{\rm 0,kpc}\;N_{\rm
cloud,21}^{-0.5}$, where $\dot{p}_{34}$ is the wind momentum flux in
10$^{34}$ dynes, $\Omega_W$ is the solid angle of the wind in
steradians, $r_{\rm 0,kpc}$ is the initial position of the cloud in
kpc, and $N_{\rm cloud,21}$ is the column density of the cloud in
10$^{21}$ cm$^{-2}$ (Strel'nitskii \& Sunyaev 1973; Heckman et
al. 2000).

A critical quantity in all of these calculations is the thermalization
efficiency, or the percentage of the mechanical energy from the
starburst that goes into heating the gas.  Unfortunately, this
quantity is poorly constrained observationally. Most simulations
assume a thermalization efficiency of 100\%, i.e. none of the energy
injected by the starburst is radiated away. In reality, this
efficiency depends critically on the environment, and is likely to be
significantly less than 100\% in the high-density environment of
powerful nuclear starbursts (e.g., Thornton et al. 1998; Strickland \&
Stevens 2000; Silich, Tenorio-Tagle, \& Mu\~{n}oz-Tu\~{n}\'on 2003).
Galactic magnetic fields may also ``cushion'' the effects of the
starburst on the ISM, and reduce the impact of the galactic wind on
the host galaxy and its environment (e.g., Tomisaka 1990; Ferri\`ere
et al. 1991; Slavin \& Cox 1992; Mineshinge et al. 1993; Ferri\`ere
1998).

\section{Observed Properties of Galactic Winds}

A great number of surveys have provided important statistical
information on galactic winds in the local universe (e.g., Heckman,
Armus, \& Miley 1990; Veilleux et al. 1995; Lehnert \& Heckman 1995,
1996; 
%Meurer et al. 1995, 1997; 
Gonzalez Delgado et al. 1998; Heckman et al. 2000, Rupke,
Veilleux, \& Sanders 2002, 2003, in prep.).  Galaxy-scale winds are
common among galaxies with global star formation rates per unit area
$\Sigma_* \equiv SFR / \pi R_{\rm opt}^2 \ga 0.1$ M$_\odot$
yr$^{-1}$ kpc$^{-2}$, where $R_{\rm opt}$ is the optical radius. This
general rule-of-thumb also appears to apply to ultra/luminous infrared
galaxies (see \S 3.3) and distant Lyman break galaxies (see \S
3.4). ``Quiescent'' galaxies with global star formation rates per unit
area below this threshold often show signs of galactic fountaining in
the forms of warm, ionized extraplanar material a few kpc above or
below the galactic disks (e.g., Miller \& Veilleux 2003a, 2003b and
references therein). The energy input from stellar winds and
supernovae in these objects elevates some of the ISM above the disk
plane, but is not sufficient to produce large-scale winds.  

This rule-of-thumb is conservative since a number of known wind
galaxies, including our own Galaxy (\S 3.1) and several dwarf
galaxies, have $\Sigma_* <<$ 0.1 M$_\odot$ yr$^{-1}$ kpc$^{-2}$ (e.g.,
Hunter \& Gallagher 1990, 1997; Meurer et al. 1992; Marlowe et
al. 1995; Kunth et al. 1998; Martin 1998, 1999; see Kunth's and
Martin's contributions at this conference). The production of
detectable winds probably depends not only on the characteristics of
the starburst (global {\em and} local $\Sigma_*$, starburst age), but
also on the detailed properties of the ISM in the host galaxies (e.g.,
see the theoretical blowout criterion of MacLow \& McCray 1988).
% and MacLow, McCray, \& Norman 1989).

The winds in actively star-forming galaxies in the local universe show
a very broad range of properties, with opening angles of $\sim$ 0.1 --
0.5 $\times$ (4$\pi$ sr), radii ranging from $<$ 1 kpc to several 10s
of kpc, outflow velocities of a few 10s of km s$^{-1}$ to more than
1000 km s$^{-1}$ (with clear evidence for a positive correlation with
the temperature of the gas phase), total (kinetic and thermal) outflow
energies of $\sim$ 10$^{53}$ -- 10$^{57}$ ergs or 5 -- 20\% of the
total mechanical energy returned to the ISM by the starburst, and mass
outflow rates ranging from $<$ 1 M$_\odot$ yr$^{-1}$ to $>$ 100
M$_\odot$ and scaling roughly with the star formation rates (see \S
3.3 below).

In the remainder of this section, we discuss a few well-studied cases
of galactic winds in the local universe and summarize the evidence for
winds in luminous and ultraluminous infrared galaxies at low and
moderate redshifts as well as in distant Lyman break galaxies. 

\subsection{The Milky Way}

By far the closest case of a large-scale outflow is the wind in our
own Galaxy. Evidence for a dusty bipolar wind extending $\sim$ 350 pc
($\sim$ 1$^\circ$) above and below the disk of our Galaxy has recently
been reported by Bland-Hawthorn \& Cohen (2003) based on data from
the Midcourse Space Experiment (MSX). The position of the warm dust
structure coincides closely with the well-known Galactic Center Lobe
detected at radio wavelengths (e.g., Sofue 2000 and references
therein). Simple arguments suggest that the energy requirement for
this structure is of order $\sim$ 10$^{55}$ ergs with a dynamical time
scale of $\sim$ 1 Myr.

Bland-Hawthorn \& Cohen (2003) also argue that the North Polar Spur, a
thermal X-ray/radio loop that extends from the Galactic plane to $b =
+80^\circ$ (e.g., Sofue 2000), can naturally be explained as an
open-ended bipolar wind, when viewed in projection in the near
field. This structure extends on a scale of 10 -- 20 kpc and implies
an energy requirement of $\sim$ (1 -- 30) $\times$ 10$^{55}$ ergs and
a dynamical timescale of $\sim$ 15 Myr, i.e.  considerably longer than
that of the smaller structure seen in the MSX maps.  If confirmed,
this may indicate that the Milky Way Galaxy has gone through multiple
galactic wind episodes.  Bland-Hawthorn \& Cohen (2003) point out that
the North Polar Spur would escape detection in external galaxies; it
is therefore possible that the number of galaxies with large-scale
winds has been (severely?) underestimated.

\subsection{Nearby Starburst Galaxies}

Two classic examples of starburst-driven outflows are described in
this section to illustrate the wide variety of processes taking place
in these objects.

\vskip 0.1in
\noindent{\bf M~82.}  This archetype starburst galaxy hosts arguably
the best studied galactic wind. Some of the strongest evidence for the
wind is found at optical wavelengths, where long-slit and Fabry-Perot
spectroscopy of the warm ionized filaments above and below the disk
shows line splittings of up to $\sim$ 250 km s$^{-1}$, corresponding
to deprojected velocities of order 525 -- 655 km s$^{-1}$ (e.g.,
%McCarthy, Heckman, \& van Breugel 1987; Bland \& Tully 1988; 
McKeith et al. 1995; Shopbell \& Bland-Hawthorn 1998). Combining these
velocities with estimates for the ionized masses of the outflowing
filamentary complex, the kinetic energy involved in the warm ionized
outflow is of order $\sim$ 2 $\times$ 10$^{55}$ ergs or $\sim$ 1\% of
the total mechanical energy input from the starburst.  The ionized
filaments are found to lie on the surface of cones with relatively
narrow opening angles ($\sim$ 5 -- 25$^\circ$) slightly tilted ($\sim$
5 -- 15$^\circ$) with respect to the spin axis of the galaxy. Deep
narrow-band images of M82 have shown that the outflow extends out to
at least 12 kpc on one side (e.g., Devine \& Bally 1999), coincident
with X-ray emitting material seen by $ROSAT$ (Lehnert, Heckman, \&
Weaver 1999) and $XMM$-Newton (Stevens, Read, \& Bravo-Guerrero
2003). The wind fluid in this object has apparently been detected by
both $CXO$ (Griffiths et al. 2000) and $XMM$-Newton (Stevens et
al. 2003). The well-known H~I complex around this system (e.g., Yun et
al. 1994) may be taking part, and perhaps even focussing, the outflow
on scales of a few kpc (Stevens et al. 2003). Recently published
high-quality CO maps of this object now indicate that some of the
molecular material in this system is also involved in the large-scale
outflow (Walter, Weiss, \& Scoville 2002; see also Garcia-Burillo et
al. 2001). The outflow velocities derived from the CO data ($\sim$ 100
km s$^{-1}$ on average) are considerably lower than the velocities of
the warm ionized gas, but the mass involved in the molecular outflow
is substantially larger ($\sim$ 3 x 10$^8$ M$_\odot$), implying a
kinetic energy ($\sim$ 3 $\times$ 10$^{55}$ ergs) that is comparable
if not larger than that involved in the warm ionized filaments. The
molecular gas is clearly a very important dynamical component of this
outflow.

\vskip 0.1in
\noindent{\bf NGC~3079. } An outstanding example of starburst-driven
superbubble is present in the edge-on disk galaxy, NGC~3079.
High-resolution $HST$ H$\alpha$ maps of this object show that the
bubble is made of four separate bundles of ionized filaments (Cecil et
al. 2001). The two-dimensional velocity field of the ionized bubble
material derived from Fabry-Perot data (Veilleux et al. 1994)
indicates that the ionized bubble material is entrained in a mushroom
vortex above the disk with velocities of up to $\sim$ 1500 km s$^{-1}$
(Cecil et al. 2001). A recently published X-ray map obtained with the
$CXO$ (Cecil, Bland-Hawthorn, \& Veilleux 2002) reveals excellent
spatial correlation between the hot X-ray emitting gas and the warm
optical line-emitting material of the bubble, suggesting that the
X-rays are being emitted either as upstream, standoff bow shocks or by
cooling at cloud/wind conductive interfaces. This good spatial
correlation between the hot and warm gas phases appears to be common
in galactic winds (Strickland et al. 2000, 2002; Veilleux et al. 2003,
and references therein). The total energy involved in the outflow of
NGC~3079 appears to be slightly smaller than that in M~82, although it
is a lower limit since the total extent of the X-ray emitting material
beyond the nuclear bubble of NGC~3079 is not well constrained (Cecil
et al. 2002). Contrary to M~82, the hot wind fluid that drives the
outflow in NGC~3079 has not yet been detected, and evidence for
entrained molecular gas is sparse and controversial (e.g., Irwin \&
Sofue 1992; Baan \& Irwin 1995; Israel et al. 1998; but see Koda et
al. 2002).

\subsection{Luminous and Ultraluminous Infrared Galaxies.}

Given that the far-infrared energy output of a (dusty) galaxy is a
direct measure of its star formation rate, it is not surprising {\em a
posteriori} to find evidence for large-scale galactic winds in several
luminous and ultraluminous infrared galaxies (LIRGs and ULIRGs; e.g.,
Heckman et al. 1990; Veilleux et al. 1995).  Systematic searches for
winds have been carried out in recent years in these objects to look
for the unambiguous wind signature of blueshifted absorbing material
in front of the continuum source (Heckman et al. 2000; Rupke et al.
2002). The feature of choice to search for outflowing neutral material
in galaxies of moderate redshifts ($z \la$ 0.6) is the Na~ID
interstellar absorption doublet at 5890, 5896 \AA. The wind detection
frequency derived from a set of 44 starburst-dominated LIRGs and
ULIRGs is high, of order $\sim$ 70 -- 80\% (Rupke et al. 2002, 2003 in
prep.; also see Rupke's and Martin's contributions at this
conference). The outflow velocities reach values in excess of 1700 km
s$^{-1}$ (even more extreme velocities are found in some AGN-dominated
ULIRGs).

A simple model of a mass-conserving free wind (details of the model
are given in Rupke et al. 2002) is used to infer mass outflow rates in
the range $\dot{M}_{\mathrm{tot}}$(H)$\;= {\mathrm few} - 120\;$ for
galaxies hosting a wind.  These values of $\dot{M}_{\mathrm{tot}}$,
normalized to the corresponding global star formation rates inferred
from infrared luminosities, are in the range $\eta \equiv
\dot{M}_{\mathrm{tot}} / \mathrm{SFR} = 0.01 - 1$. The parameter
$\eta$, often called the ``mass entrainment efficiency'' or
``reheating efficiency'' shows no dependence on the mass of the host
(parameterized by host galaxy kinematics and absolute $R$- and
$K^{\prime}$-band magnitudes), but there is a possible tendency for
$\eta$ to decrease with increasing infrared luminosities (i.e. star
formation rates). The large molecular gas content in ULIRGs may impede
the formation of large-scale winds and reduce $\eta$ in these objects.
A lower thermalization efficiency (i.e. higher radiative efficiency)
in these dense gas-rich systems may also help explain the lower $\eta$
(Rupke et al. 2003, in prep.; see Rupke's contribution at this
conference).

\subsection{Lyman Break Galaxies}

Evidence for galactic winds has now been found in a number of z $\sim$
3 -- 5 galaxies, including an important fraction of Lyman break
galaxies (LBGs; e.g., Franx et al. 1997; Pettini et al. 2000, 2002;
Frye, Broadhurst, \& Benitez 2002; Dawson et al. 2002; Ajiki et
al. 2002; Adelberger et al. 2003; Shapley et al. 2003). The best
studied wind at high redshift is that of the gravitationally lensed
LBG MS~1512-cB58 (Pettini et al. 2000, 2002). An outflow velocity of
$\sim$ 255 km s$^{-1}$ is derived in this object, based on the
positions of the low-ionization absorption lines relative to the
rest-frame optical emission lines (Ly$\alpha$ is to be avoided for
this purpose since resonant scattering and selective dust absorption
of the Ly$\alpha$ photons may severely distort the profile of this
line; e.g., Tenorio-Tagle et al. 1999). The mass-conserving free wind
model of Rupke et al. (2002) applied to MS~1512-cB58 (for consistency)
results in a mass outflow rate of $\sim$ 20 $M_\odot$ yr$^{-1}$,
equivalent to about 50\% the star formation rate of this galaxy based
on the dust-corrected UV continuum level. Similar outflow velocities
are derived in other LBGs (Pettini et al. 2001). The possibly strong
impact of these LBG winds on the environment at high $z$ is discussed
in the next section (\S 4.2).

\section{Implications}

Starburst-driven winds may have a strong influence on structure formation
at high redshifts, on the ``porosity'' of star-forming galaxies
(i.e. the probability for ionizing photons to escape their host
galaxies), hence the nature of the extragalactic UV and infrared
background, and on the chemical and thermal evolution of galaxies and
their environment.  Due to space limitations, this review only
addresses the last issue.

\subsection{Heating and Enrichment of the ISM and IGM}

\noindent{\bf Hot Metal-Enriched Gas.}  Nuclear starbursts inject both
mechanical energy and metals in the centers of galaxies. This hot,
chemically-enriched material, the driving engine of galactic winds, is
eventually deposited on the outskirts of the host galaxies, and
contributes to the heating and metal enrichment of galaxy halos and
the IGM. Surprisingly little evidence exists for the presence of this
enriched wind fluid. This is due to the fact that the wind fluid is
tenuous and hot and therefore very hard to detect in the X-rays. The
current best evidence for the existence of the wind fluid is found in
M~82 (Griffiths et al. 2000; Stevens et al. 2003), NGC~1569 (Martin,
Kobulnicky, \& Heckman 2002), and possibly the Milky Way (e.g., Koyama
et al. 1989; Yamauchi et al. 1990). The ratio of alpha elements to
iron appears to be slightly super-solar in the winds of both NGC~1569
and M~82, as expected if the stellar ejecta from SNe II are providing
some, but not all of the wind fluid.

\vskip 0.1in
\noindent{\bf Selective Loss of Metals. }  The outflow velocities in
LIRGs and ULIRGs do not appear to be correlated with the rotation
velocity (or equivalently, the escape velocity) of the host galaxy,
implying selective loss of metal-enriched gas from shallower
potentials (Heckman et al. 2000; Rupke et al. 2002).  If confirmed
over a broader range of galaxy masses (e.g., Martin 1999; but see the
contribution by Martin at this conference for a word of warning), this
result may help explain the mass-metallicity relation and radial
metallicity gradients in elliptical galaxies and galaxy bulges and
disks (e.g., Bender, Burstein, \& Faber 1993; Franx \& Illingworth
1990; Carollo \& Danziger 1994; Zaritsky et al. 1994; Trager et
al. 1998).
%; Jablonka et al. 1996; Jorgensen et al. 1996; Pahre et al. 1998;
%Several models have assumed that this was the case (e.g., Wyse \& Silk
%1985; Dekel \& Silk 1986; Lynden-Bell 1992; Kauffmann \& Charlot 1998
%and many others since).
The ejected gas may also contribute to the heating and chemical
enrichment of the ICM in galaxy clusters (e.g., Dupke \& Arnaud 2001;
Finoguenov et al. 2002, and references therein).
%Finoguenov, Arnaud, \& David 2001; Tamura et al. 2001

\vskip 0.1in
\noindent{\bf Dust Outflows.}  Galactic winds also act as conveyor
belts for the dust in the hosts. The evidence for a large-scale dusty
outflow in our own Galaxy has already been mentioned in \S 3.1
(Bland-Hawthorn \& Cohen 2003). Far-infrared maps of external galaxies
with known galactic winds show extended dust emission along the galaxy
minor axis, suggestive of dust entrainment in the outflow (e.g., Hughes,
Gear, \& Robson 1994; Alton et al. 1998, 1999;
%; Alton, Davies, \& Bianchi 1999; 
Radovich, Kahanp\"a\"a, \& Lemke 2001).
% NGC 253 H_2: Sugai et al. 2003). 
Direct evidence is also found at optical wavelengths in the form of
elevated dust filaments in a few galaxies (e.g., NGC~1808, Phillips
1993; NGC~3079, Cecil et al. 2001).
%NGC~891, Howk \& Savage 1997).
A strong correlation between color excesses, $E(B - V)$, and the
equivalent widths of the blueshifted low-ionization lines in
star-forming galaxies at low (e.g., Armus, Heckman, \& Miley 1989;
Veilleux et al. 1995; Heckman et al. 2000; Rupke et al. 2003) and
moderate-to-high redshifts (e.g., Rupke et al. 2003; Shapley et
al. 2003) provides additional support for the prevalence of dust
outflows. Assuming a Galactic dust-to-gas ratio, Heckman et al.
(2000) estimate that the dust outflow rate is about 1\% of the total
mass outflow rate in LIRGs. Dust ejected from galaxies may help feed
the reservoir of intergalactic dust (e.g., Coma cluster; Stickel et
al. 1998).
%; note, however, that tidal and ram-pressure stripping is
%probably more efficient than winds at carrying dust into the ICM; see
%contribution by Stickel at this conference).

\subsection{Zone of Influence of Winds}

The impact of galactic winds on the host galaxies and the environment
depends sensitively on the size of the ``zone of influence'' of these
winds, i.e.  the region affected either directly (e.g., heating,
metals) or indirectly (e.g., ionizing radiation) by these winds. This
section summarizes the methods used to estimate this quantity.

\vskip 0.1in
\noindent{\bf Indirect Measurements based on Estimates of the Escape
Velocity.}  The true extent of galactic winds is difficult to
determine in practice due to the steeply declining density profile of
both the wind material and the host ISM. The zone of influence of
galactic winds is therefore often estimated using indirect means which
rely on a number of assumptions.  A popular method is to use the
measured velocity of the outflow and compare it with the local escape
velocity derived from some model for the gravitational potential of
the host galaxy. If the measured outflow velocity exceeds the
predicted escape velocity {\em and} if the halo drag is negligible, then
the outflowing material is presumed to escape the host galaxy and be
deposited in the IGM on scales $\ga$ 50 -- 100 kpc. This method was
used for instance by Rupke et al. (2002) to estimate the average
escape fraction $\langle f_{\mathrm{esc}} \rangle \equiv \sum
\dot{M}_{\mathrm{esc}}^i / \sum \dot{M}_{\mathrm{tot}}^i$ and
``ejection efficiency'' $\langle\delta\rangle \equiv \sum
\dot{M}_{\mathrm{esc}}^i / \sum \mathrm{SFR}^i$ for 12 ULIRGs, which
were found to be $\sim 0.4-0.5$ and $\sim 0.1$, respectively.  These
calculations assumed that the host galaxy could be modeled as a singular
isothermal sphere truncated at some radius $r_{\rm max}$. Neither the
escape fraction nor the ejection efficiency were found to be sensitive
to the exact value of $r_{\rm max}$. Other strong cases for escaping
material include the ``H$\alpha$ cap'' of M~82 and the $\sim$ 1500 km
s$^{-1}$ line-emitting material in the superbubble of NGC~3079; both
objects were discussed in \S 3.2.

Note that the outflow velocities measured by Rupke et al. (2002) refer
to the neutral component of the outflow, not the hot enriched wind
fluid.  Unfortunately, direct measurements of the wind velocity are
not yet technically possible so one generally relies on the expected
terminal velocity of an adiabatic wind at the measured X-ray
temperature $T_X$ [$v_X \sim (5KT_X/\mu)^{0.5}$, where $\mu$ is the
mean mass per particle] to provide a lower limit to the velocity of
the wind fluid (this is a lower limit because it only takes into
account the thermal energy of this gas and neglects any bulk motion;
e.g., Chevalier \& Clegg 1985; Martin 1999; Heckman et al. 2000).
%This method has been used recently by Martin et al. (2002) to argue
%that an important fraction of the wind fluid detected in NGC~1569 is
%likely to escape this galaxy...

Arguably the single most important assumption made to determine the
fate of the outflowing gas is that halo drag is negligible.  Silich \&
Tenorio-Tagle (2001) have argued that halo drag may severely limit the
extent of the wind and the escape fraction.  Drag by a dense halo or a
complex of tidal debris may be particularly important in ULIRGs if
they are created by galaxy interactions (e.g., Veilleux, Kim, \&
Sanders 2002b).

\vskip 0.1in
\noindent{\bf Deep X-ray and Optical Maps of Local Starbursts.}  The
fundamental limitation in directly measuring the zone of influence of
winds is the sensitivity of the instruments. Fortunately, $CXO$ and
$XMM$-Newton now provide powerful tools to better constrain the
extent of the hot medium (e.g., M~82, Stevens et al. 2003; NGC~3079,
Cecil et al. 2002; NGC~6240, Komossa et al. 2003; Veilleux et
al. 2003; NGC~1511, Dahlem et 2003). The reader should refer to the
contribution of M. Ehle at this conference for a summary of recent
X-ray results (see also Strickland et al. 2003 and references
therein).

The present discussion focusses on optical constraints derived from
the detection of warm ionized gas on the outskirts of wind
hosts. Progress in this area of research has been possible thanks to
advances in the fabrication of low-order Fabry-Perot etalons which are
used as tunable filters to provide monochromatic images over a large
fraction of the field of view of the imager. The central wavelength
(3500 \AA\ -- 1.0 $\mu$m) is tuned to the emission-line feature of
interest and the bandwidth (10 -- 100 \AA) is chosen to minimize the
sky background. Continuum and emission-line images are produced nearly
simultaneously thanks to a ``charge shuffling/frequency switching''
mode, where the charges are moved up and down within the detector at
the same time as switching between two discrete frequencies with the
tunable filter, therefore averaging out temporal variations associated
with atmospheric lines and transparency, seeing, instrument and
detector instabilities.  The narrow-band images are obtained in a
straddle mode, where the off-band image is made up of a pair of images
that ``straddle'' the on-band image in wavelength (e.g., $\lambda_1$ =
6500 \AA\ and $\lambda_2$ = 6625 \AA\ for rest-frame H$\alpha$); this
greatly improves the accuracy of the continuum removal since it
corrects for slopes in the continuum and underlying absorption
features.

These techniques have been used with the Taurus Tunable Filter (TTF;
Bland-Hawthorn \& Jones 1998; Bland-Hawthorn \& Kedziora-Chudczer
2003) on the AAT and WHT to produce emission-line images of several
``quiescent'' disk galaxies (Miller \& Veilleux 2003a) and a few
starburst galaxies (Veilleux et al. 2003) down to unprecedented
emission-line fluxes. Gaseous complexes or filaments larger than
$\sim$ 20 kpc have been discovered or confirmed in a number of wind
hosts (e.g., NGC~1482 and NGC~6240; the presence of warm ionized gas
at $\sim$ 12 kpc from the center of M~82 was discussed in \S 3.2).
Multi-line imaging and long-slit spectroscopy of the gas found on
large scale reveal line ratios which are generally not H~II
region-like.  Shocks often contribute significantly to the ionization
of the outflowing gas on the outskirts of starburst galaxies. As
expected from shock models (e.g., Dopita \& Sutherland 1995), the
importance of shocks over photoionization by OB stars appears to scale
with the velocity of the outflowing gas (e.g., NGC~1482, NGC~6240, or
ESO484-G036 versus NGC~1705; NGC~3079 is an extreme example of a
shock-excited wind nebula; Veilleux et al. 1994), although other
factors like the starburst age, star formation rate, and the dynamical
state of the outflowing structure (e.g., pre- or post-blowout) must
also be important in determining the excitation properties of the gas
at these large radii (e.g., Shopbell \& Bland-Hawthorn 1998 and
Veilleux~\&~Rupke~2002).

\vskip 0.1in
\noindent{\bf Influence of the Wind on Companion Galaxies.}  Companion
galaxies located within the zone of influence of the wind will be
affected by the wind ram pressure. Irwin et al. (1987) noticed that
the dwarf S0 galaxy NGC~3073 exhibits an elongated H~I tail that is
remarkably aligned with the nucleus of NGC~3079. Irwin et al. have
argued that ram pressure due to the outflowing gas of NGC~3079 is
responsible for this tail. If that is the case, the wind of NGC~3079
must extent to at least $\sim$ 50 kpc. This is the only system known
so far where this phenomenon is suspected to take place.

\vskip 0.1in
\noindent{\bf Absorption-Line Studies.}  Absorption-line spectroscopy
of bright background galaxies (e.g., high-$z$ quasars, Lyman break
galaxies) can provide direct constraints on the zone of influence of
galactic winds. Norman et al. (1996) have used this method to estimate
the extent of the wind in NGC~520. A strong and possibly complex Mg
II, Mg I, and Fe II absorption-line system was found near the systemic
velocity of NGC~520 at a distance from the galactic nucleus of 24
$h^{-1}$ kpc. A weaker system at a distance of 52 $h^{-1}$ kpc is also
possibly present. Unfortunately, NGC~520 is undergoing a tidal
interaction so the absorption may arise from tidally disrupted gas
rather than material in the purported wind. Norman et al. also looked
for absorption-line systems associated with the wind of NGC~253, but
the proximity of this system to our own Galaxy and to the line of
sight to the Magellanic Stream makes the identification of the
absorption-line systems ambiguous. No other local wind galaxy has been
studied using this technique.

%other Mg II absorption line studies: Bergeron et al 1992, 1994, 
%Srianand \& Khare 1994, Lanzetta et al. 1995

Large absorption-line data sets collected on high-$z$ galaxies provide
new constraints on the zone of influence of winds in the early
universe. Adelberger et al. (2003) have recently presented tantalizing
evidence for a deficit of neutral hydrogen clouds within a comoving
radius of $\sim$ 0.5 $h^{-1}$ Mpc from $z \sim 3$ LBGs. The
uncertainties are large and the results are significant at
less than the $\sim$ 2$\sigma$ level. Adelberger et al. (2003)
argue that this deficit, if real, is unlikely to be due solely to the
ionizing radiation from LBGs (e.g., Steidel et al. 2001; Giallongo et
al. 2002). They favor a scenario in which the winds in LBGs directly
influence the surrounding IGM. They also argue that the excess of
absorption-line systems with large CIV column densities near LBGs is
evidence for chemical enrichment of the IGM by the LBG winds.

\section{Summary}

Tremendous progress has been made over the past five years in
understanding the physics and impact of starburst-driven winds in the
local and distant universe. We now know that starburst-driven winds
are common in galaxies with high star formation rates per unit area,
both locally (nearby starbursts, luminous and ultraluminous infrared
galaxies) and at high redshifts (e.g., Lyman break galaxies). There is
{\em direct} evidence that starburst-driven winds have had a strong
influence on the chemical evolution of the host ISM and possibly also
that of the IGM: (1) Enriched wind fluid has been detected in a few
nearby galaxies. (2) Approximately half of the outflowing material in
powerful starburst galaxies (ULIRGs) have velocities in excess of the
escape velocities. (3) Deep emission-line maps of local wind galaxies
indicate that the zone of influence of the wind often extends beyond
$\sim$ 10 kpc. (4) Recent results on Lyman break galaxies suggest
tentatively that the zone of influence of LBG winds may extent out to
$\sim$ 500 $h^{-1}$ kpc. Nevertheless, much work remains to be done in
this area of research; the next five years promise to be equally
exciting as the last five!

\acknowledgements

It is a pleasure to thank S. Aalto for organizing a stimulating
conference.  Some of the results presented in this paper are part of a
long-term effort involving many collaborators, including
J. Bland-Hawthorn, G. Cecil, P. L. Shopbell, and R. B. Tully and
Maryland graduate students S. T. Miller and D. S. Rupke. This article
was written while the author was on sabbatical at the California
Institute of Technology and the Observatories of the Carnegie
Institution of Washington; the author thanks both of these
institutions for their hospitality.  The author acknowledges partial
support of this research by a Cottrell Scholarship awarded by the
Research Corporation, NASA/LTSA grant NAG 56547, and NSF/CAREER grant
AST-9874973.

\end{document}